\documentclass[twocolumn,showpacs,preprintnumbers,amsmath,amssymb]{revtex4}
                                                                                
\usepackage{graphicx}
\usepackage{dcolumn}
\usepackage{bm}
                                                                                
\begin{document}
                                                                                
\preprint{APS/123-QED}
                                                                                
\title{Surface reconstruction and ferroelectricity in PbTiO$_3$ thin films }

\author{M.Sepliarsky, M.G.Stachiotti and R.L.Migoni}

\address{Instituto de F\'{\i}sica Rosario, Universidad Nacional de Rosario,  
         27 de Febrero 210 Bis, (2000) Rosario, Argentina \\ \\}

\begin{abstract}
Surface and ferroelectric properties of PbTiO$_3$ thin films are investigated
using an interatomic potential approach 
with  parameters computed from first-principles calculations.
We show that a model developed for the bulk describes properly  
the surface properties of PbTiO$_3$. In particular, 
the antiferrodistortive surface reconstruction, recently observed 
from X-ray scattering, is correctly reproduced as 
a result of the change in the balance of long-range Coulombic 
and short-range interactions at the surface.
The effects of the surface reconstruction
on the ferroelectric properties of ultrathin films are 
investigated.
Under the imposed open-circuit electrical boundary conditions, 
the model gives a critical thickness for ferroelectricity of 
4 unit cells. The surface layer, which forms the antiferrodistortive 
reconstruction, participates in the ferroelectricity. 
A decrease in the tetragonality of the films leads to the stabilization of a phase 
with non-vanishing in-plane polarization.
A peculiar effect of the surface reconstruction 
on the in-plane polarization profile is found.

\end{abstract}

\maketitle


\section{Introduction}

One fascinating feature of perovskites is that they exhibit a large variety 
of structural phase transitions. 
The variety of transition behavior and low-temperature distorted
structures depend on the individual compound.
Among the perovskites one finds ferroelectric (FE) crystals such as 
BaTiO$_3$, KNbO$_3$ (displaying three phase transitions), and 
PbTiO$_3$ (displaying only one transition), antiferroelectrics such as PbZrO$_3$, 
and materials such as SrTiO$_3$ that exhibit other non-polar 
antiferrodistortive (AFD) instability involving the rotation of the oxygen 
octahedra~\cite{lines}. 

With the rapidly advancing miniaturization of ferroelectric devices and the use of
thin films, attention is focusing on the role played 
by surfaces and interfaces in the overall performance of the materials.   
The effects of surfaces on the structural phase transitions have become an issue 
of significant importance because the surface can affect the structural 
behavior of the perovskites by modifying the strength of various instabilities. 
Lead titanate is a clear example. In PbTiO$_3$ (PT), ferroelectricity is 
due to the condensation of a $\Gamma_{15}$
unstable phonon in which the oxygen octahedra shift against the Pb sub-lattice.
The ground state consists of shifts along (001) accompanied by a tetragonal lattice
strain which stabilizes this direction.~\cite{coh92} 
However, complete phonon dispersions for the ideal perovskite
structure PbTiO$_3$ as determined by density functional calculations show  
ferroelectric ($\Gamma_{15}$) and also rotational antiferrodistotive ($R_{25}$ type) 
instabilities.~\cite{gho99} In bulk, the FE and AFD instabilities compete with each 
other and the FE lattice distortion suppresses the AFD distortion, but 
the proximity to a surface modifies that balance. 
Recently, an antiferrodistortive reconstruction of the PbTiO$_3$ (001) surface 
has been found
using grazing incidence X-ray scattering. The atomic structure of the 
surface consist of a single layer of an AFD structure with oxygen
cages rotated by 10$^o$ around the [001] axis through the Ti ions.~\cite{mun02} 
Latter on, the AFD 
reconstruction at the PbO-terminated (001) surface was reproduced by Bungaro 
and Rabe using ab-initio calculations.~\cite{bun04} 
They also found for in-plane polarized films
that the FE and AFD distortions coexist in the proximity of the surface.
Regarding device applications, out-of-plane polarized films are more relevant.
Therefore, it is worth investigating also the effects of the surface
reconstruction on the out-of-plane ferroelectricity, since 
it was speculated that the AFD reconstruction may be related to the dead layer 
often postulated to explain the observed decrease in ferroelectric character 
in thin films.~\cite{mun02}     

The realistic simulation of PT thin-films is a theoretical challenge due to the  
interplay of polar and non-polar instabilities at the surface. 
Although first-principles methods are extremely
precise and include thoroughly electronic effects,  
they are quite computer demanding. 
So, these calculations are restricted to 
the investigation of zero-temperature properties of perovskites involving a 
rather small number of atoms. For the study of the thermal behavior, or 
to 
handle larger system sizes, other methods are necessary. 
During the last years, the effective Hamiltonian approach has been used
for the investigation of thin film properties~\cite{gho00,wu04,kor04,alma04}. 
However, for the 
simulation of Pb-based perovskite thin-films the coexistence of rotational and FE
distortions should be explicitly considered.~\cite{for01}   
Atomistic simulations based on interatomic potentials can  
account naturally  for the presence of competing instabilities.
However, the validity of any atomistic simulation study depends to 
considerable extent on the quality of the potentials used.
The interatomic potential approach firmly grounded 
by having its parameters fitted against results of first-principles calculations 
is a promising technique for investigating thin-film properties 
of Pb-based perovskites.
In this paper we investigate the ferroelectric properties of PT thin films
using a shell model developed completely from first-principles calculations.
We investigate first if the model developed for the bulk describes properly  
surface properties. Then the ferroelectric behavior
of ultrathin films under open-circuit electrical boundary conditions are 
investigated.

\section{Model and computational details}

The approach we follow in this work is based on the 
atomistic modeling using interatomic potentials.
We indeed found that the shell model approach does provide 
computationally efficient and confident methodology for the simulation of 
ferroelectric perovskites, 
including bulk properties of pure crystals~\cite{sep99,tin99,sep04}
and solid solutions~\cite{sep00}, superlattices~\cite{sep01}, surface 
and thin films properties.~\cite{tin01}  
In the shell model, each 
atom is represented by a massive core coupled
to a massless shell, and the relative core-shell displacement 
describes the atomic polarization. The model contains coulombic long-range  
and pairwise short-range interactions. 
In the present study we use the model developed in Ref ~\cite{sep04}
which contains 4$^{th}$ order core-shell 
couplings and short-range interactions described by two 
different types of potentials. 
A Rydberg potential ($V(r)=(a+br)\exp(-r/\rho )$)
is used for the Pb-Ti, Pb-O and Ti-O pairs, and a Buckingham potential
($V(r)=a\exp(-r/\rho)+cr^{-6}$) is used for O-O interactions.
The model was developed completely from first-principles calculations
within the Local Density Approximation (LDA).
The parameters were adjusted to reproduce a wide number of LDA
results, including lattice dynamics and total energy calculations.
The resulting model was able to reproduce delicate properties of PT in
good agreement with LDA results. However, as a consequence of the LDA, 
volume and volume dependent properties are underestimated
with respect to experimental values.
For example, the model gives a tetragonal ground state with a lattice 
parameter a=3.859$\AA$, a tetragonal distortion c/a=1.043, and a 
spontaneous polarization P=$54 \mu$C/cm$^2$, while the experimental data are 
a=3.90$\AA$, c/a=1.065, and P=$75 \mu$C/cm$^2$ .
The underestimation of the static structural properties is translated via the adjusted model 
to the finite
temperature behavior. Molecular dynamics simulations showed a cubic-tetragonal 
transition at T$_C$ = 450K, 300K below the experimental value.
Nevertheless, the qualitative temperature behavior  
of lattice parameters and polarization was correctly reproduced. 
See Reference~\cite{sep04} for more details. 

The investigation of the surface properties is carried out using an isolated slab 
geometry with periodic boundary conditions in the x-y plane. 
The long ranged electrostatic energy and forces are calculated 
by a direct sum method~\cite{wolf}.
The equilibrated zero-temperature structure of the slabs was determined by a 
zero-temperature quench until the force on each individual ion was 
less than 0.001 eV/$\AA$. To mimic the two dimensional clamping and 
straining of the film due to the presence of a substrate, 
we force the simulation cell to be square in the x-y plane.

\section{Results and discussion}
\subsection{Surface effects in non-polar slabs. Validation of the model}

The parameters of the model were derived to describe perfect crystal 
properties, and it is not guaranteed that the same potentials are suitable to
describe the surface properties of PT.
In the bulk, the underlying potential surface is a result
of a delicate balance between
long range Coulomb, short range, and core-shell interactions.
Such balance could change significantly 
at the surface, where the atomic environment changes due to
the discontinuity of the crystal.
So, it is important to check if the model developed for the bulk   
proves also successful for describing surface properties.

A first-principles study for (001) surfaces of cubic PT has been
performed by Meyer, Padilla and Vanderbilt using $c(1\times1)$ surface 
periodicity ~\cite{meyer99}. As a first step, we take these first-principles 
calculations as benchmark results to compare with. To this end we determined 
the atomic equilibrium position for both PbO- and TiO- symmetrically terminated 
7-layer periodic slabs. 
We set the in-plane lattice parameter to the cubic equilibrium value yielded by
the model in bulk, a=3.887\AA, and we relaxed the atomic positions from the
ideal positions only in the direction perpendicular to the surface.
The results obtained for the atomic relaxations for both slabs are listed in 
Table ~\ref{tab1}, whereas the values for the interlayer distances and
rumplings are listed in Table ~\ref{tab2}. Ab-initio results 
are shown in parenthesis as reference. 
The model reproduces satisfactorily the relaxation direction of the
atoms, the change in the interlayer distances, and atomic rumplings.
For the case of the PbO surface, the equilibrium termination 
for PT~\cite{meyer99}, the numerical values are in very good agreement with 
the ab-initio ones. The average surface relaxation energy
obtained by the model $\triangle E_{relax}=0.237 eV$ is also in quite good
agreement with the ab-initio result of 0.210 eV.

\begin{table}
\begin{tabular}{cc|ccc}
\hline
  Atom        &  $\delta_z$   &   & Atom         & $\delta_z$    \\   
\hline
 Pb(1)        & -3.45 (-4.36) &   & Ti(1)        & -4.17 (-3.40) \\
 O$_{III}$(1) & -0.27 (-0.46) &   & O$_{I}$(1)   & -2.94 (-0.34) \\
 Ti(2)        &  3.19 ( 2.39) &   & O$_{II}$(1)  & -2.94 (-0.34) \\
 O$_{I}$(2)   &  1.65 ( 1.21) &   & Pb(2)        &  1.58 ( 4.53) \\
 O$_{II}$ (2) &  1.65 ( 1.21) &   & O$_{III}$(2) & -0.34 ( 0.43) \\
 Pb(3)        & -0.78 (-1.37) &   & Ti(3)        & -0.98 (-0.92) \\
 O$_{III}$(3) &  0.33 (-0.20) &   & O$_{I}$(3)   &  0.56 (-0.27) \\
              &               &   & O$_{II}$(3)  &  0.56 (-0.27) \\
\hline
\end{tabular}
\caption{Atomic relaxations perpendicular to the surface ($\delta_z$) of the PbO (left panel) 
and the TiO$_2$ (right panel) terminated surface in the cubic phase. $\delta_z$
are given as percent of the theoretical unit cell parameter a. 
For comparison, ab-initio results (Ref~\cite{meyer99}) 
are shown in parentheses.}
\label{tab1}
\end{table}

\begin{table}
\begin{tabular}{cc|c}
\hline
              &  PbO surface  &   TiO surface   \\
\hline
$ \Delta d_{12}$     &  -4.04 (-4.2) &  -4.17 (-4.4)   \\
$ \Delta d_{23}$     &   2.34 ( 2.6) &  +0.83 (+3.1)   \\
$ \Delta d_{34}$     &  -0.02 (-0.8) &  -0.21 (-0.6)   \\
$\nu_{1}$     &   3.02 (3.9)  &   1.22 (3.1)    \\
$\nu_{2}$     &   1.85 (1.2)  &   1.93 (4.1)    \\
$\nu_{3}$     &   0.96 (1.2)  &   1.54 (0.7)    \\

\end{tabular}
\caption{Change in the interlayer distance ($\Delta d$) and layer rumpling ($\nu$) 
for the PbO (left panel) and the TiO$_2$ (right panel)
terminated surface in the cubic phase, given as percent of the
theoretical unit cell parameter a.
For comparison, ab-initio results (Ref~\cite{meyer99})
are shown in parentheses.}

\label{tab2}
\end{table}

The recent pseudopotential study of the AFD surface reconstruction in
PbTiO$_3$ slabs~\cite{bun04} provides reliable additional
information to validate the model.
In Table ~\ref{tab3} we show structural parameters of the relaxed $c(2\times2)$
11-layer PbO terminated slab. The overall agreement with the ab-initio results 
is very good.
The model reproduces the main differences between the
$c(2\times2)$ and $(1\times1)$ surface periodicities, that is the presence 
of an AFD reconstruction with a reduction of rumplings and interlayer distances. 
Regarding the AFD surface reconstruction, the rotation angles are slightly overestimated 
with respect to the pseudopotential calculations; the experimentally determined 
rotation angle for the surface layer is 10$^o$~\cite{mun02}. The energy gain associated
with the octahedra rotations is 0.32 eV per surface unit cell, which is 
approximately one order of magnitude larger than the bulk ferroelectric well depth.     

\begin{table}
\begin{tabular}{cc|cc|cc}
\hline
Interlayer & & Rumpling & & Angle & \\
\hline
$ \Delta d_{12}$ & -3.3 (-3.4) & $\nu_{1}$ & -1.4 (-1.4) & $\theta_{2}$ &  13.8 (11.4) \\
$ \Delta d_{23}$ & +2.1 (+2.9) & $\nu_{2}$ & +1.3 (+0.9) & $\theta_{4}$ &  -1.7 (-2.9)  \\
$ \Delta d_{34}$ & -0.1 (-0.9) & $\nu_{3}$ & -1.4 (-2.0) & $\theta_{6}$ &   8.0 (3.9)   \\
$ \Delta d_{45}$ & +0.3 (+0.4) & $\nu_{4}$ & +0.5 (+0.4) & $\theta_{bulk}$ &6.0 (3.3)  \\
$ \Delta d_{56}$ & +0.1 (-0.1) & $\nu_{5}$ & +0.2 (-0.2) &               &    \\
\end{tabular}
\caption{Change in the interlayer distance ($\Delta d$), layer rumpling ($\nu$)
and rotation angle of oxygen octahedra ($\theta$) for the PbO terminated 
surface at $a_{bulk}$ in a 11-layer non-polar PbTiO$_3$ slab. 
For comparison, ab-initio results (Ref.~\cite{bun04})
are shown in parentheses.}
                                                            
\label{tab3}
\end{table}

It was pointed out that the enhancement of the AFD distortion at the 
PbO surface is a consequence of the particular chemistry of Pb, due to its tendency 
to move off-center and form strong covalent Pb-O bonds~\cite{bun04}. 
We have demonstrated that this electronic 
effect, which has been showed to be an important factor for ferroelectricity 
in PbTiO$_3$~\cite{coh92}, can be mimicked at the atomic level by a shell model.
In this less-sophisticated approach, the AFD surface reconstruction is 
obtained just as a result of the change in the balance of long-range Coulombic and 
short-range interactions at the surface.

\subsection{Ferroelectric behavior in polar slabs}

We use the atomistic model to explore zero-temperature ferroelectric properties in 
PT ultrathin films. We consider only PbO-terminated
surfaces because they are the most stable ones.~\cite{meyer99}
The simulation cell contains $10\times10$ unit cells 
with periodic boundary conditions along the $x-y$ plane. 
The in-plane lattice parameter was set and clamped to the model equilibrium
value for the  
tetragonal bulk a=3.859$\AA$. Since this value is close to the 
lattice parameter of SrTiO$_3$, this 
clamped condition  simulates the strain effect of 
a SrTiO$_3$ substrate.

We have performed standard-atomic relaxation methods to determine
the zero temperature structure in slabs from 2 to 10 unit cells thick
that contains from 1200 to 5200 atoms, respectively.
In each case, the relaxation started with atoms slightly displaced 
from their ideal cubic positions. After a first quenching, the system
is iteratively warmed-up and quenched again in order to 
eliminate possible metastable configurations. 
To analyze the results, we define the cell parameter $c$ as the 
distance between two consecutive PbO planes, and 
the local polarization is defined as the polarization of a Ti-centered 
unit cell.

A decrease in tetragonality is an important effect
produced by the presence of the surface.
Figure 1 provides detailed microscopic information about 
the average cell-by-cell tetragonal distortion $c/a$ for a 10-unit 
cells thick slab. 
It can be seen that the tetragonality of all unit cells in the slab is
considerable lower than in the bulk ($c/a=1.043$ for the bulk).
The stronger reduction is observed for the surface cells 
which are practically cubic ($c/a=1.005$) while 
the tetragonality increases towards the interior of the slab
reaching the value $c/a=1.032$.  
As the imposed in-plane lattice parameter corresponds with the one of 
ferroelectric bulk PT, we can expect that the tetragonality gradually 
converges to the bulk value when the film thickness increases. 
In fact, we have obtained that, while the tetragonality of the 
outermost layer practically does not depend on film thickness (surface relaxation effects
are predominant for these layers), the inner cells display a thickness dependence.    
Figure 2 shows the evolution of the c/a ratio with film thickness for cells 
in the middle of the slab. 
We can roughly estimate from the extrapolation of Figure 2 that $d=86 \AA$ ($\approx$ 
22 unit cells) is the 
lower-bond estimate of the slab thickness requiered to reach the tetragonality of the bulk
in the center of the film. 
The strong thickness dependence of tetragonality is in qualitative agreement with recent 
experimental measurements which showed that the average c/a ratio decreases substantially 
for films thinner than $200 \AA$ ~\cite{lich04}.

\begin{figure}[tbp] 
\centering
 \includegraphics[height=2.8in,width=3.8in]{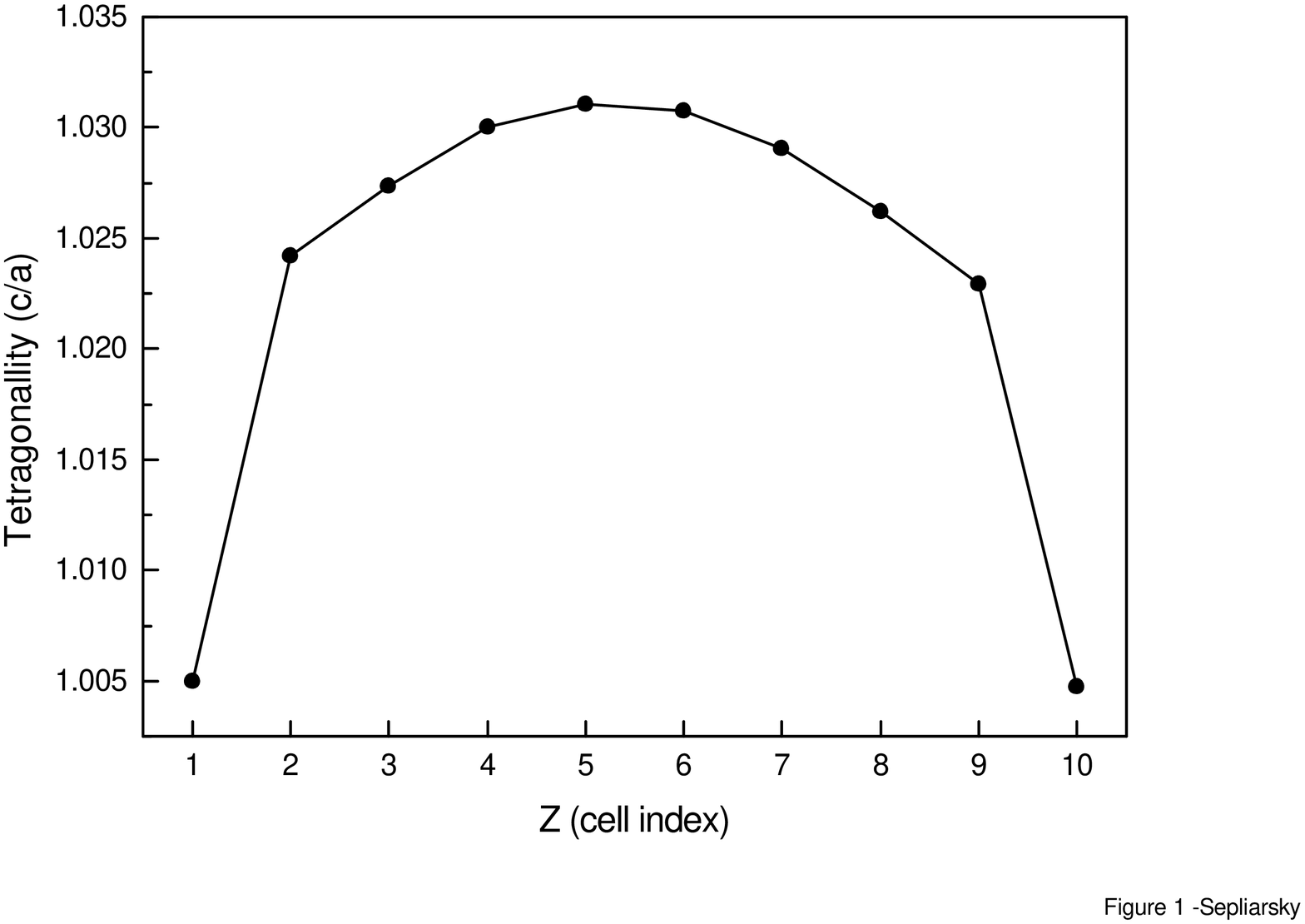}
  \caption{Average cell-by-cell tetragonal distortion $c/a$ for a 
           10-unit cells thick slab.}
  \label{fig1}
\end{figure}

\begin{figure}[t] 
\centering
 \includegraphics[height=3.3in,width=3.5in]{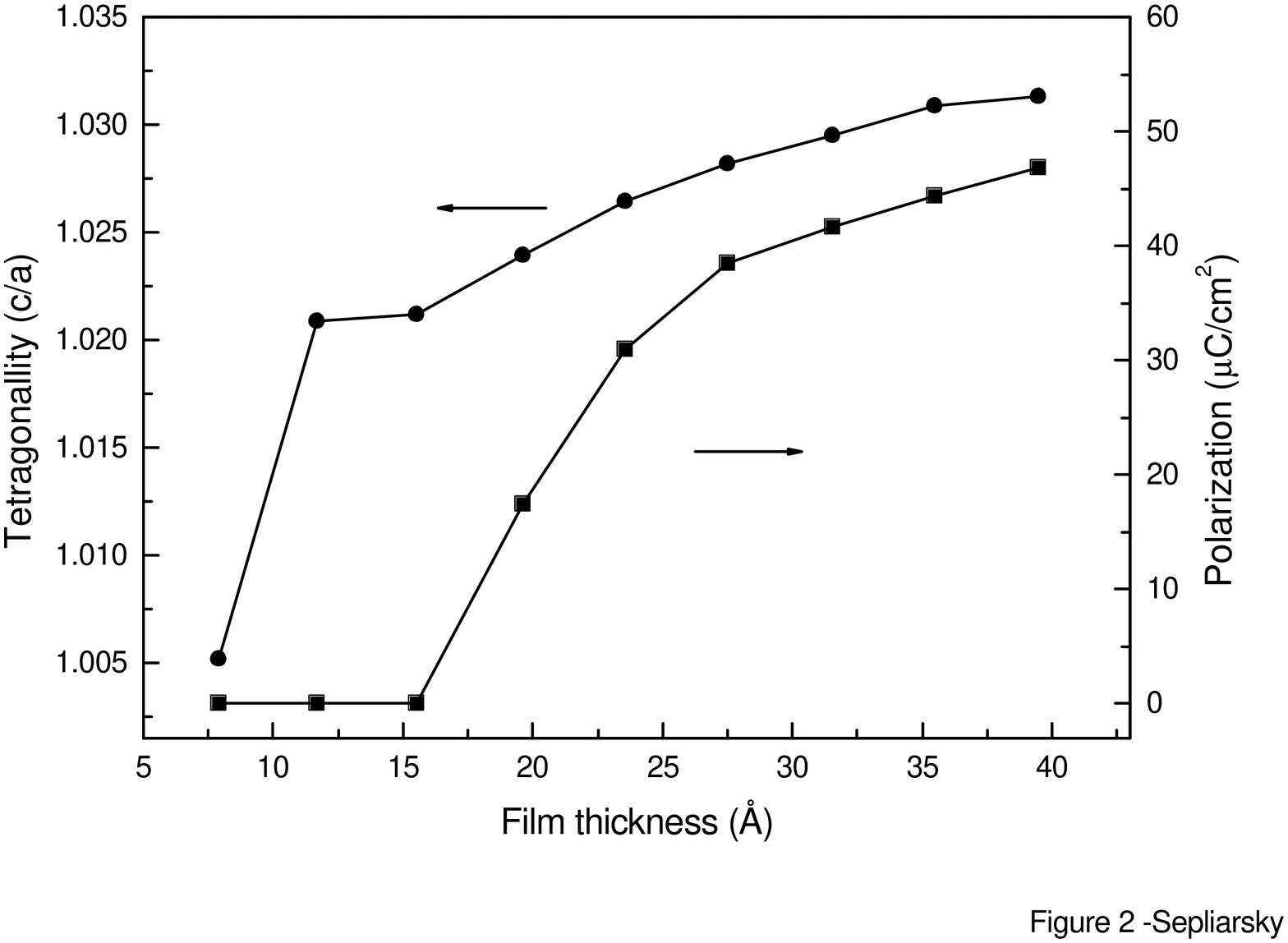}
  \caption{Thickness dependence of the tetragonal distortion (for the
           inner cells) and the spontaneous out-of-plane polarization of
           the ferroelectric domains.}
  \label{fig2}
\end{figure}
                                                                                
In slabs under ideal open-circuit electrical boundary conditions, as we are considering here, 
out-of-plane ferroelectricity manifests above a critical thickness, through the formation 
of stripe domains of alternating polarity. These domains are effective in 
neutralizing the depolarized field and stabilizing the ferroelectric phase, even without
electrodes.\cite{tin01,wu04,kor04}
The observation of stripe domains below the ferroelectric transition in epitaxial PbTiO$_3$ ultrathin films on SrTiO$_3$ substrates was reported from X-ray scattering ~\cite{stre02}, 
and the ferroelectric phase was found to be stable for thickness down to 3 
unit cells~\cite{fong04}.
Figure 3 shows the cell-by-cell out-of-plane polarization 
profile of a chain perpendicular to the surface of slabs of 4 and 5 unit 
cell width. It is clear that under the imposed stress and electrical boundary conditions, 
the model gives a critical thickness ($d_c$) for ferroelectricity of 4 unit cells ($d_c$ 
is defined as the maximun thickness at which polarization is zero).
In the 4-cell slab the surface cells develop a small inward polarization 
of $\approx 3 \mu C/cm^2$ due to surface atomic relaxations (see Table~\ref{tab1}) while 
the polarization of the two inner cells is practically zero. 
In the 5-cell slab, individual chains have a net out-of-plane polarization 
but the average polarization of the slab is zero due to the development of stripe 
domains (the evolution of the polarization at the center of the slab 
with film thickness is shown in Figure 2). 
The inner cells display a local polarization of 
$\approx 18 \mu C/cm^2 $, that is 0.33 of the bulk value. 
The surface cells present an even stronger reduction of polarization.
It is very important to point out, however, that the  
polarization of the surface cells in the 5-cell 
ferroelectric slab is considerable larger than in the 4-cell  
non-ferroelectric one, it 
increases from  $\approx 3 \mu C/cm^2$ to $\approx 10 \mu C/cm^2$.
This polarization enhancement indicates that 
the surface layer, which forms an AFD reconstruction, also participates in the 
ferroelectricity. 
We note that the value of $d_c$ obtained with the model is two unit cells larger 
than the experimental one reported in Ref.~\cite{fong04}. 
This difference could arise from the fact the model was developed to reproduce the LDA 
ground state of bulk PT, and consecuently the ferroelectric instability is underestimated.
Another possibility is that we are not explicity simulating the presence of a 
substrate (although we are considering its strain effects) which modifies the  
boundary condition at one face of the film.  
Atomistic investigations of the substrate effects
on the critical thickness are in progress.   
   
In the simulations, the in-plane lattice parameter was set to the tetragonal 
bulk value a=3.859$\AA$, so a vanishing 
in-plane polarization component could be expected. However, this is not the case and 
the slabs develop a polarization component along the (110) direction. 
For the bulk, the coupling of the ferroelectric soft mode with strain is 
the main driving force for the stabilization of the (001)-polarized tetragonal phase over 
the (111)-polarized rhombohedral one~\cite{coh92}. In films, the decrease in 
tetragonality favors the stabilization of a phase 
with a non-vanishing in-plane polarization component, above and also below the 
critical thickness. 
Figure 4 shows the change of the average in-plane polarization with film thickness.
The reduction of tetragonality produces that the in-plane polarization increases 
when the film thickness decreases. 
Locally, a strong increment of the in-plane polarization 
at the surface is observed (see the profile in the inset of Figure 4). 
The polarization decreases towards the interior of the 
slab. 
For thick enough films, it is expected that the interior of the slab reachs 
the vanishing in-plane polarization of the bulk.

\begin{figure}[tbp] 
\centering
 \includegraphics[height=3.3in,width=3.8in]{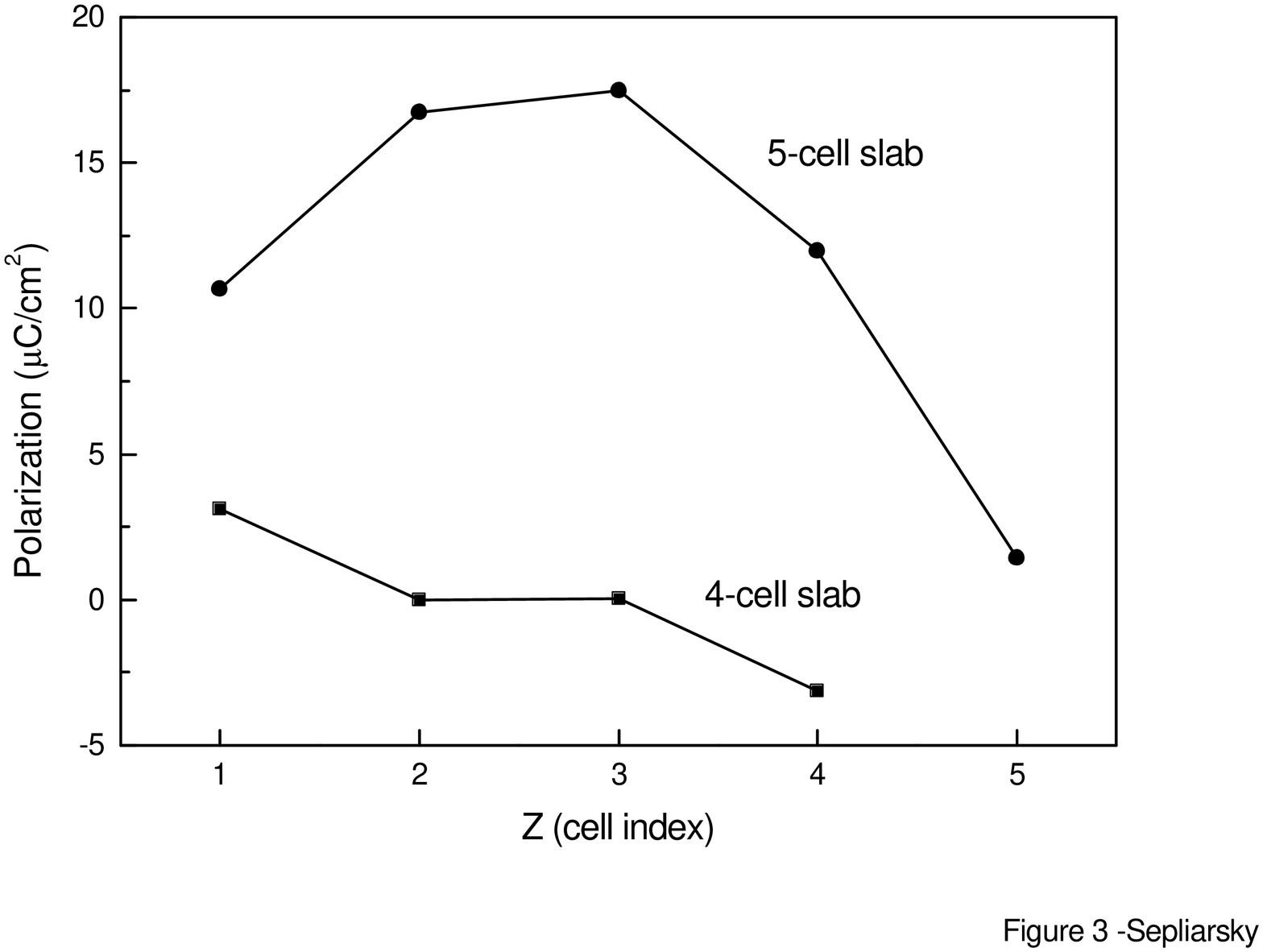}
  \caption{Cell-by-cell out-of-plane polarization profile of
           a randomly chosen chain perpendicular to the surface
           for a slab of 4 and 5 unit cell width.}
  \label{fig3}
\end{figure}

We showed above that non-polar slabs displayed an AFD surface reconstruction.
Polar slabs show a surface reconstruction with a similar    
rotation angle which is independent on the slab thickness.
However, one difference is that
neighboring planes of octahedra along the rotation axis rotates in-phase instead
of out-of-phase. In fact, for the polar slabs, $\theta_{2}=13.5^o$, 
$\theta_{4}=4.5^o$ and $\theta_{6}=7.3^o$ for the first, second and 
third layers, respectively. We note that unstable modes at the M and R points, which are 
associated with oxygen rotation instabilities, have similar frequencies and compete 
to each other in the cubic bulk~\cite{gho99}. For the bulk, the model yields to an 
energy difference of only 2 meV per unit cell between the M and R distortions
($\Delta$ E = 1.2 meV per unit cell from ab-initio calculations~\cite{bun04}). 
The presence of the polarized surface affects these two competing structural 
instabilities stabilizing M-type distortions at the surface.

Finally, we show a peculiar effect of the AFD surface reconstruction 
on the in-plane polarization profile. 
Although the average in-plane polarization is oriented along the (110) 
direction, unit cells are not uniformly polarized along that direction.
This can be visualized by the top view of the surface showed in Figure 5.  
We will consider the displacements of the Pb atoms to make the description 
simpler. It is clear from the figure that the Pb atoms displace mainly along (100) 
directions (see the arrows) in such a way that half of the Pb atoms are 
displaced along a (100) direction, and the other half along a (010) 
direction. 
Although the effect is much stronger at the surface, this local behavior
is observed throughout the ultrathin film.
This polarization profile, which produces an average in-plane polarization along 
the (110) direction, is a consequence of the oxygen octahedra rotation plus the 
tendency of Pb to move off-center shortening Pb-O bonds.

\begin{figure}[tbp] 
\centering
 \includegraphics[height=3.3in,width=3.8in]{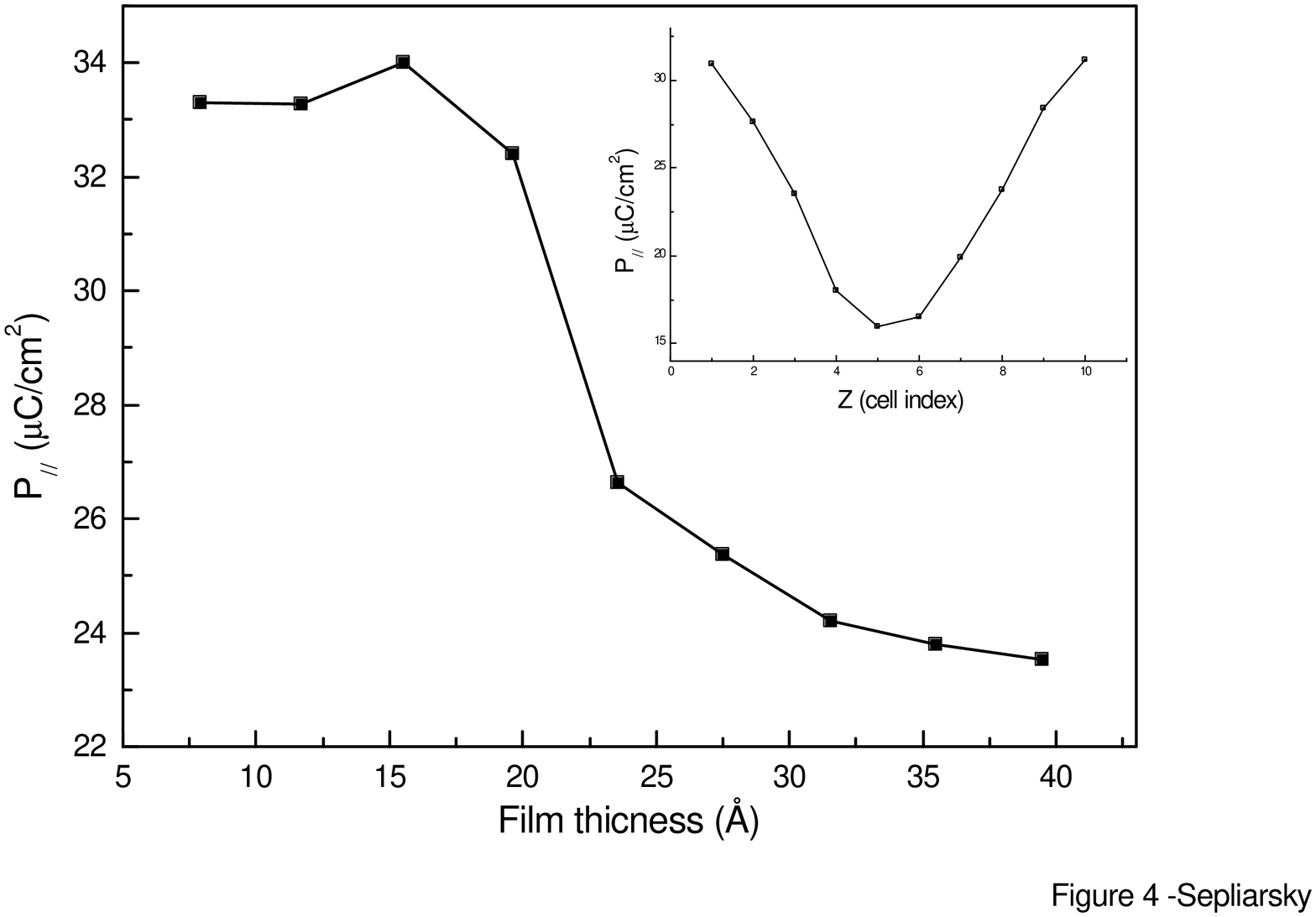}
  \caption{Evolution of the average in-plane polarization with film thickness.
           Inset: cell-by-cell in-plane polarization profile for
           a randomly chosen chain perpendicular to the surface
           in a slab of 10 unit cell width.}
  \label{fig4}
\end{figure}

In summary, we have shown that the delicate balance of lattice strain and polar  
and non-polar instabilities, which is responsible for the observed
tetragonal ground-state of bulk PbTiO$_3$ and the AFD structure of the 
surface, is accurately simulated by an interatomic potential approach with  
parameters computed from first-principles calculations. 
In ultrathin films, a remarkable decrease in tetragonality
leads to the stabilization of a phase with non-vanishing 
in-plane polarization, above and also below the critical thickness ($d_c$) for 
ferroelectricity. 
Under the imposed stress and electrical boundary conditions,
$d_c$ was found to be 4 unit cells. 
The surface layer, which forms an AFD reconstruction, also participates in the 
ferroelectricity. 

\begin{figure}[tbp] 
\centering
 \includegraphics[height=3.0in,width=3.8in]{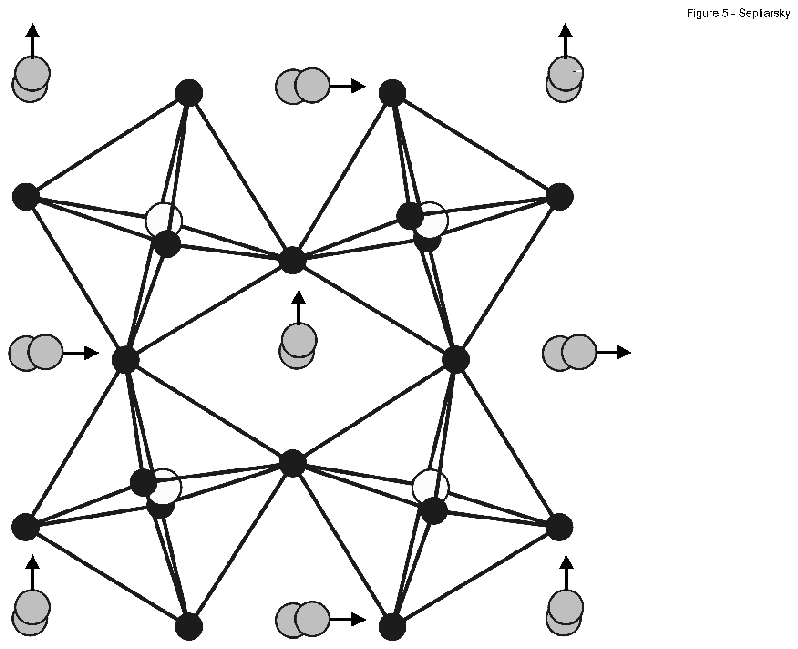}
  \caption{Top view of the antiferrodistortive surface reconstruction.
           The arrows indicate the displacements of the Pb atoms which 
           produce an average in-plane polarization oriented along 
           the (110) direction.}
  \label{fig5}
\end{figure}

\acknowledgments
This work was supported by Agencia Nacional de Promoci\'on Cient\'{\i}fica y Tecnol\'ogica, and 
Consejo Nacional de Investigaciones Cientif\'{\i}cas y T\'ecnicas (Argentina). 
M.G.S. thanks support from Consejo de Investigaciones de la Universidad Nacional de 
Rosario. \\ \\

\end{document}